\documentclass[conference]{IEEEtran}

\IEEEoverridecommandlockouts 

\usepackage{cite}
\usepackage[cmex10]{amsmath}
\usepackage{amssymb}
\usepackage{bm}
\usepackage{tikz,pgfplots} 
\usetikzlibrary{plotmarks}
\usepackage{flushend}

\usepackage{url}
\usepackage[hidelinks]{hyperref}
\usepackage{multirow}
\usepackage{booktabs,colortbl}
\usepackage{todonotes}

\newcommand{\tcb}[1]{\textcolor{blue}{#1}}
\newcommand{\tcg}[1]{\textcolor{black}{#1}}

\AtEndDocument{\par\leavevmode}

\begin{document}
\title{
%
Monitoring and Prediction in Smart Energy Systems
via Multi-timescale Nexting
}

\author{%

\IEEEauthorblockN{Johannes Feldmaier, Dominik Meyer, Hao Shen, Klaus Diepold}
\IEEEauthorblockA{Department of Electrical and Computer Engineering,
Technische Universit\"at M\"unchen, Germany \vspace*{-2em}
}


}
\maketitle

\begin{abstract}
Reliable prediction of system status is a highly demanded
functionality of smart energy systems, which can enable users or
human operators to react quickly to potential future system changes.
By adopting the multi-timescale nexting method, we develop an
architecture of human-in-the-loop energy control system, which is
capable of casting short-term predictive information about the 
specific smart energy system.
The developed architecture does either require a system model nor 
additional acquisition of (sensor) data in the existing system configuration. 
%
Our first experiments demonstrate the performance of the proposed 
control architecture in an electrical heating system simulation. 
In the second experiment, we verify the effectiveness of our developed 
structure in simulating a heating system in a thermal model of a building,
by employing natural \textit{EnergyPlus} temperature data.
\end{abstract}


\section{Introduction}
%
%
Continuous monitoring of energy systems is crucial for ensuring reliability and 
security of the systems, such as generator, grid infrastructure,
and residential home. 
There has been a significant effort in developing failure detection techniques 
and outlier detection algorithms. 
These methods are expected to deliver reliable estimations or predictions 
of the system status, so as to assist human operators (or users) with 
early warnings or updates of potential issues. 
Commonly, statistics or system models are employed to detect anomalies in 
the monitored (sensor) data, cf. \cite{smart2015hayes, 
failure2011calderaro, collins2012smart, sequeira2014energy, linda2012comp}.

Most predictive control systems use one timescale to predict the future behaviour of 
the considered system. This means the predictions about what is to be happening
in the future is restricted to a fixed number of seconds or timesteps.
On the other side, human beings as well as other animals 
seem to use experiences from earlier situations to anticipate what is about to 
happen next and adjust their actions accordingly. 
Those living things which have been capable of making accurate predictions 
about the future are better prepared for suitable actions and perceptions than others. 
This ability makes it easier for them to take advantage of upcoming opportunities 
as well as to evade future danger. The process of continuously anticipating the 
immediate future in a local and personal sense is called \textit{nexting}. 

The work in \cite{nexting2014modayil} demonstrates a technical implementation 
of the nexting behavior on a mobile robot. The robot was able to learn how to 
simultaneously predict all its raw sensor signals at different timescales in 
real time. In a recent work, such predictions of raw sensor signals are used in 
a laser welding robot to improve the quality of the weld seam by adjusting the 
process parameters adaptively \cite{gunt:mach15}. 
A similar concept of automata learning has been also applied to model, analyze,
and detect anomalies in energy consumption of a system, cf. \cite{gilani2013importance}.

In this work, we propose an architecture of human-in-the-loop energy control 
system, which is capable of predicting semantically meaningful information 
about an energy system. It enables human operators to take such predictive 
knowledge into account to further adjust the goal (or system configuration) 
to keep the system in an optimal state. 
Our proposed architecture consists of three interative parts, namely,
the physical energy system to be monitored, the operator (or user), 
and the \textit{NEXTMon} system. 
The actual physical system is monitored constantly, and 
grants access to its system states (sensor readings, statistical data) and 
controller actions. These data are processed by the NEXTMon system and displayed 
to the user as additional predictive information. The operator as the 
human-in-the loop assesses all available data and controls the 
physical system. This human-in-the-loop architecture is depicted in 
Figure~\ref{fig:nextmon}.

Compared to conventional prediction algorithms the NEXTMon system does not 
require an exact system model and therefore is not limited to specific energy 
systems. The NEXTMon system combines external information sources (like weather 
forecasts), sensor measurements, and control actions in order to learn the system 
behavior. While conventional algorithms often rely on fixed models or regression 
functions, our approach is able to approximate arbitrary functions (similar to neural 
networks) and adapt them to changing conditions. Learning a predictive model 
generally requires a lot of data samples and only predicts one timescale. The 
proposed nexting algorithm requires only few data samples per update step and 
simultaneously updates weight vectors for multiple timescales.

In the following, we describe both tile coding (Section \ref{sec:tilecoding}) and 
the nexting algorithm (Section \ref{sec:nexting}) as they are the main components 
of the NEXTMon system. In Section \ref{sec:application} and \ref{sec:experiment} we 
describe an example how the system can be used in a room heating scenario. Finally, 
we present some experimental results in Section~\ref{sec:result} and a conclusion in
Section~\ref{sec:conclusion}.

\section{State representation as multi sensor observations}
\label{sec:tilecoding}
Reinforcement Learning (RL) \cite{sutton1998rl} is an important machine 
learning discipline and has been successfully applied to solve model free control
problems. 
A common task of RL is to learn the so-called value function which is
designed or constructed to reflect the specific control task, and 
often defined as the expected reward. 
The reward signal often carries incomplete information towards the 
ultimate goal of the problem. Designing the reward signal differently enables 
applications of RL in robotics, control, and economics. 

A most classic RL method is the so-called temporal difference (TD) 
learning algorithm. Arguably, the most convenient characteristic of  
TD methods is via calculating the difference in estimates of the value 
function between simply two consecutive time steps. Computationally, 
this requires even less effort than the common stochastic gradient algorithms.
%
%
This data efficiency is convenient in domains where the current state is 
acquired with lot of sensors. The algorithm applied in this work is a 
combination of the basic reinforcement learning framework with a temporal 
difference update for making short term predictions
(c.f. Section \ref{sec:nexting}).

Working with RL requires a unique and extensive system representation incorperating 
all available system information.  
In small, simulated domains, where RL has proven as an effective learning technique, 
a good state representation is easily achievable. However, in real world problems
where the system state is represented as multiple continuous sensor readings, this could 
result in a prohibitively large, or infinite, state space.

\begin{figure}
\centering
\includegraphics[width=0.45\textwidth]{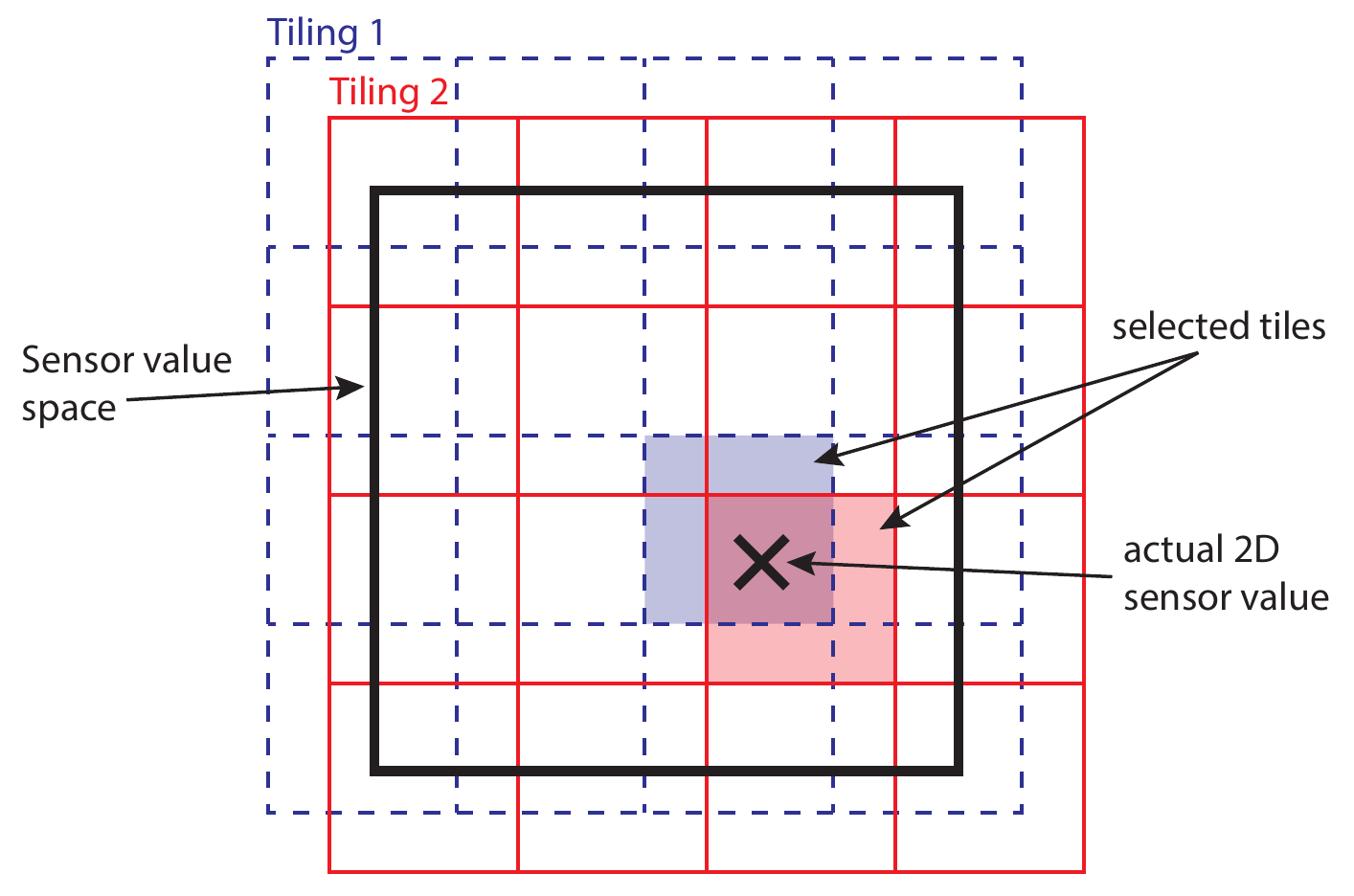}
\caption{Tile coding with two tilings for a 2D sensor value state 
space (adopted from \cite{sutton1998rl})}
\label{fig:tile_coding}
\end{figure}

One technique to overcome this difficulty is the tile coding technique
\cite{sutton1998rl}, which achieves a good balance between accuracy of representation,
computational costs, and complexity. Tile coding is widely adopted and used in different 
disciplines of RL. 
In tile coding, the sensor value space is partitioned into \textit{tiles} 
(c.f. Figure \ref{fig:tile_coding}). The complete partition covering the whole 
sensor value space is called a \textit{tiling}. For one sensor value space 
there can exist several overlapping tilings. The algorithm determines for 
each sensor value the corresponding position within all tilings. 
The overall resolution of tile coding is determined by granularity and 
generalisation parameters. Granularity is set by the number of overlapping 
tilings and the generalization parameters describe the shape of each tile. 
Dividing a sensor value space into $4\times 4$ tiles and using two different tiles
like in the example of Figure \ref{fig:tile_coding} results in a coarse 
generalisation between sensor values, that are within $0.25$ of each others 
for both dimensions. A second tiling with an offset to the first one 
refines this generalisation. This results in an overall resolution for 
this example of $0.25/2$ or $0.125$. 

The resulting feature vector representing the actual sensor state is then 
calculated as follows. Tile coding determines the actual activated tile 
for each sensor value and each tiling. All tiles on each tiling are 
sequentially-numbered, thus for each activated tile the corresponding index 
can be determined. With all activated tiles or indexes a binary feature 
vector is created. A binary column vector with a length equal to the number 
of all available tiles is used to represent the state of activated tiles. 
Each entry of the feature vector $\bm{\phi}_t$ for each activated tile is 
set to one, all other entries are set to zero. 

\subsection*{Joint Tile Coding}

In Figure \ref{fig:tile_coding} an example of joint tile coding is 
shown. Generally tile coding can be used for each sensor reading and 
each state value independently. Each value is coded separately and 
the calculated feature vectors can be concatenated afterwards. Thus 
tile coding can be considered as a mapping from state information to 
feature representation and hence formulated as 
\begin{equation}
    \bm{\phi}_{t} = f(\bm{x}_t, a_t), 
\end{equation}
where $\bm{x}_t$ corresponds to all available sensor data, $a_t$ to 
the actions the controller has performed at time step $t$, and $f(\cdot)$ 
the non-linear mapping performed by tile coding. In this work 
we use bold face notation for vectors and matrices.

In most real world applications there are different types of sensors 
and variables representing the actual state. An efficient way to achieve 
a state representation using tile coding resulting in more distinct feature 
vector in such domains is joint tile coding. Joint tile coding 
works better than independent tile coding provided that there is limited 
interaction between the different state dimensions. So for different 
types of sensors and state values joint tile coding groups are created 
and there values are coded jointly. 

Compared to other feature selection techniques such as radial basis
functions, Kanerva coding, etc. tile coding delivers an intuitive 
way for representing the state features without requiring complex 
feature engineering.
%
%
\section{Multi-timescale predictions using Nexting}
\label{sec:nexting}
The common goal of RL is to learn the value function that computes the long-term 
expected reward. In the multi-timescale nexting setting, the raw sensor signals take 
on this role of reward within the learning algorithm, they are called \textit{pseudo
rewards}.

For each raw sensor signal several predictions with different timescales will 
be made at each discrete time step $t$. We 
indicate variables by the index $i$ to point out that the quantity corresponds to 
a prediction with the specific timescale $i$. 

The sensor reading at time $t$, i.e. the pseudo reward at time $t$ concerning 
the $i$th prediction, is denoted by $R_{t}^{i} \in \mathbb{R}$. 
The overall discounted sum of the respective future pseudo rewards $R_{t}^{i}$, denoted
by the {\it return} $G_{t}^{i}$, is defined to be the ideal prediction $V_{t}^{i} \in \mathbb{R}$:
%
\begin{equation}
  \label{eq:predictioninmultitimescalenexting}
    V_{t}^{i} \:\: := \:\: \sum_{k=0}^{\infty} (\gamma^{i})^{k} R_{t+k+1}^{i} = G_{t}^{i} \; ,
\end{equation}
where $\gamma^{i} \in [0,1)$ is the \textit{discount rate} for the $i$th 
prediction. 
Here, $G_{t}^{i}$ is the ideal value for 
the $i$th prediction at time step $t$ -- the so-called \textit{ideal prediction}.
It is worth noticing that this ideal prediction is simply an approximation of the real
sensor signal to be predicted.
In order to reflect the correct timescale of the prediction, it is crucial to 
choose an appropriate discount rate. 
Specifically, the discount rate $\gamma^{i}$ for a timescale of 
$\tau^{i}$ time steps can be determined by
\begin{equation}
  \label{eq:discountrateforatimescaleofttimesteps}
    \gamma^{i} = 1 - \frac{1}{\tau^{i}} \; .
\end{equation}
Multi-timescale nexting uses linear function approximation to compute each 
prediction. If $\bm{\phi}_{t} \in \mathbb{R}^{N}$ denotes the \textit{feature vector} 
with $N$ features characterizing the state of the 
system at time step $t$, all predictions $V_{t}^{i}$ can be generated with the 
scalar products of the feature vector $\bm{\phi}_{t}$ and the appropriate 
weight vector $\bm{\theta}_{t}^{i} \in \mathbb{R}^{N}$ denoted as
\begin{equation}
  \label{eq:prediction}
    V_{t}^{i} \approx \bm{\phi}_{t}^{\top} \bm{\theta}_{t}^{i} = \sum_{j=1}^{N} \bm{\phi}_{t,j} \,\bm{\theta}_{t,j}^{i} \; .
\end{equation}
Here $\bm{\phi}_{t}^{\top}$ is the transpose of the feature vector $\bm{\phi}_{t}$ while 
$\bm{\phi}_{t,j}$ and $\bm{\theta}_{t,j}^{i}$ denotes the $j$th component of each vector. 
The feature vector $\bm{\phi}_{t}$ is calculated using tile coding (see Section 
\ref{sec:tilecoding}). 

For learning these weight vectors, the \textit{linear gradient-descent 
TD($\lambda$)} algorithm is used. The update rule for learning the weight 
vectors $\bm{\theta}_{t}^{i}$ at each time step $t$ is
\begin{equation}
  \label{eq:multitimescalenextingupdateruleforweightvector}
    \bm{\theta}_{t+1}^{i} = \bm{\theta}_{t}^{i} + \alpha \, \delta_{t}^{i} \, \textbf{z}_{t}^{i} \; ,
\end{equation}
where $\alpha > 0$ is a step-size parameter (which influences the rate of 
learning) and 
\begin{equation}
  \label{eq:tderror}
   \delta_{t}^{i} = R_{t+1}^{i} + \gamma^{i}  \bm{\phi}_{t+1}^{\top} \bm{\theta}_{t}^{i} - \bm{\phi}_{t}^{\top} \bm{\theta}_{t}^{i}
\end{equation}
is the \textit{TD error} for the $i$th prediction at time step $t$. Furthermore, 
$\textbf{z}_{t}^{i} \in \mathbb{R}^{n}$ denotes the vector of \textit{accumulating 
eligibility traces}. Eligibility traces serve as extra memory variables which 
are linked to each state characterized by the feature vector $\bm{\phi}_{t}$. 
The initial value of the eligibility trace vector $\textbf{z}_{t}^{i}$ is 
$\textbf{0}$. Afterwards, the eligibility trace is updated in each step $t$ by 
\begin{equation}
  \label{eq:multitimescalenextingupdateruleforeligibilitytrace}
    \textbf{z}_{t}^{i} = \gamma^{i} \, \lambda \, \textbf{z}_{t-1}^{i} + \bm{\phi}_{t} \; ,
\end{equation}
where $\lambda \in [0,1]$ is called the trace-decay parameter. By Equation 
(\ref{eq:multitimescalenextingupdateruleforeligibilitytrace}), the eligibility 
trace of all currently present features is incremented by 1, whereas all other 
features, i.e. all features which are currently nonpresent, are decayed by 
$\gamma^{i} \lambda$. 

In this way, the learned weight vector represents an 
implicit knowledge about the underlying process. It is constantly updated 
with each new observation and adapts to changing conditions and stores them 
in the corresponding weights. Therefore, the interplay between number of features 
and the way they are extracted are the most crucial part of the nexting 
algorithm. Tile coding is an appropriate technique to calculate such unique 
feature vectors with sufficient entropy. 

\begin{figure}
\centering
\includegraphics[width=0.45\textwidth]{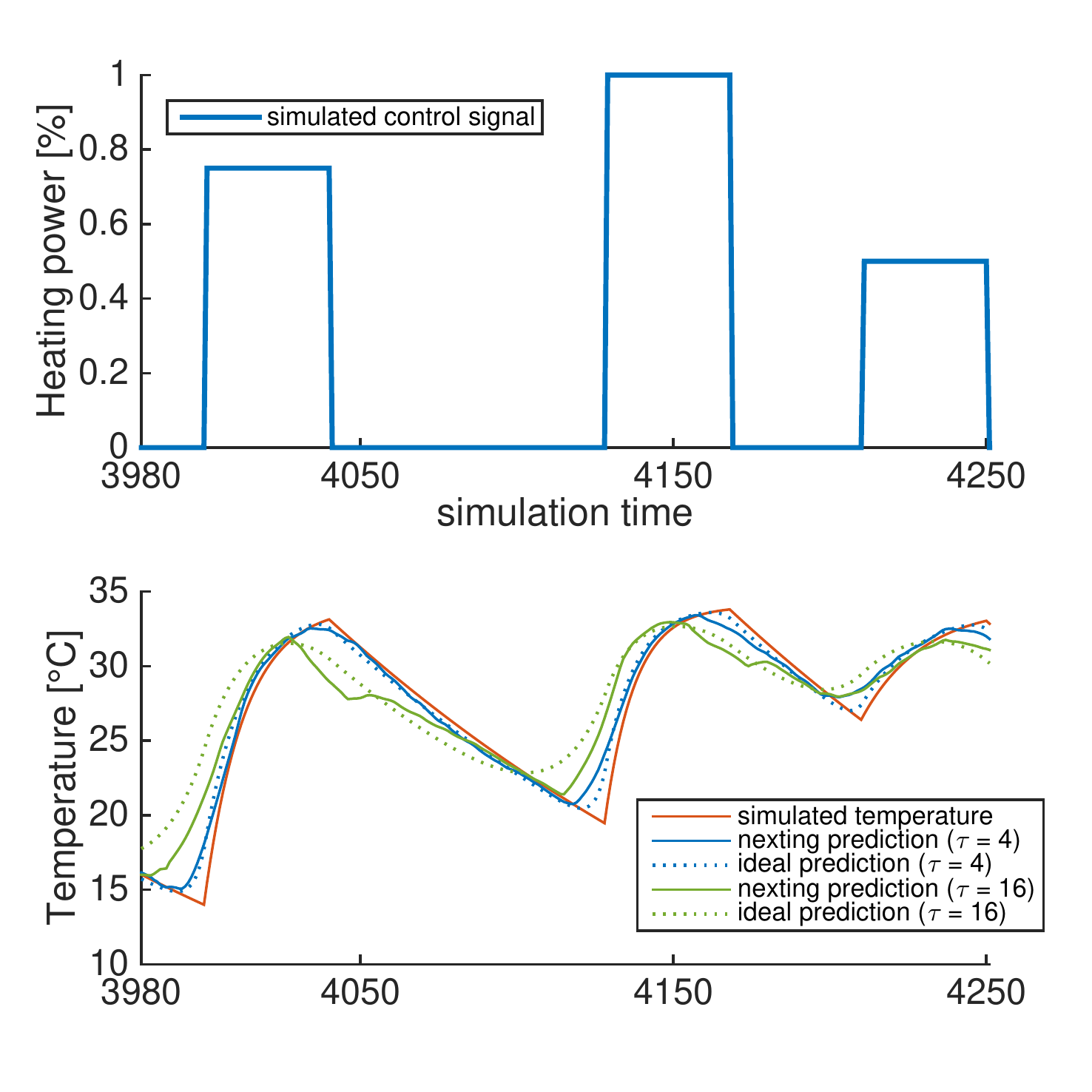}	
\caption{Example of a prediction produced by the nexting algorithm.}
\label{fig:nexting_demo}
\end{figure}

\section{Multi-timescale Nexting for Monitoring Applications}
\label{sec:application}
As described in the previous section, the nexting algorithm uses all available 
state information $\bm{\phi}_{t}$ and a weight vector $\bm{\theta}_{t}^{i}$ 
for calculating a prediction $V^i_t$. In Figure \ref{fig:nexting_demo} we 
depicted a result of the nexting algorithm. In this example we simulated the 
characteristic of heating an insulated water tank. The heater is turned on with 
different power levels (50\%, 75\%, and 100\%), the temperature (red line) of 
the water tank increases during heating and it cools down after switching 
the heater of. Firstly, we calculated the ideal prediction $G_t^i$ for 
$\gamma^1 = 0.75$ and $\gamma^2 = 0.9375$ corresponding to a prediction 
time of $\tau^1 = 4$ and $\tau^2 = 16$ time steps ahead respectively 
(dotted lines). Then we used the nexting algorithm for calculating an 
online prediction of the temperature signal. In tile coding, only the 
control and the temperature signals were used to calculate the actual 
feature vector. At each time step the feature vector is expanded with 
a history of four preceding feature vectors to add more information. 
Therefore, each update step only requires the information of the actual 
and four preceding states. After an initial learning phase (in this 
example after about 3000 time steps) the corresponding weight vectors 
for each prediction $V^i_t$ are sufficiently approximated. Multiplying 
the actual feature vector $\bm{\phi}_{t}$ with each weight vector 
$\bm{\theta}_{t}^{i}$ results in multiple predictions as plotted in 
blue (for 4 time steps ahead) and green (for 16 time steps ahead) in 
Figure \ref{fig:nexting_demo}. 

\begin{figure}
\centering
\includegraphics[width=0.45\textwidth]{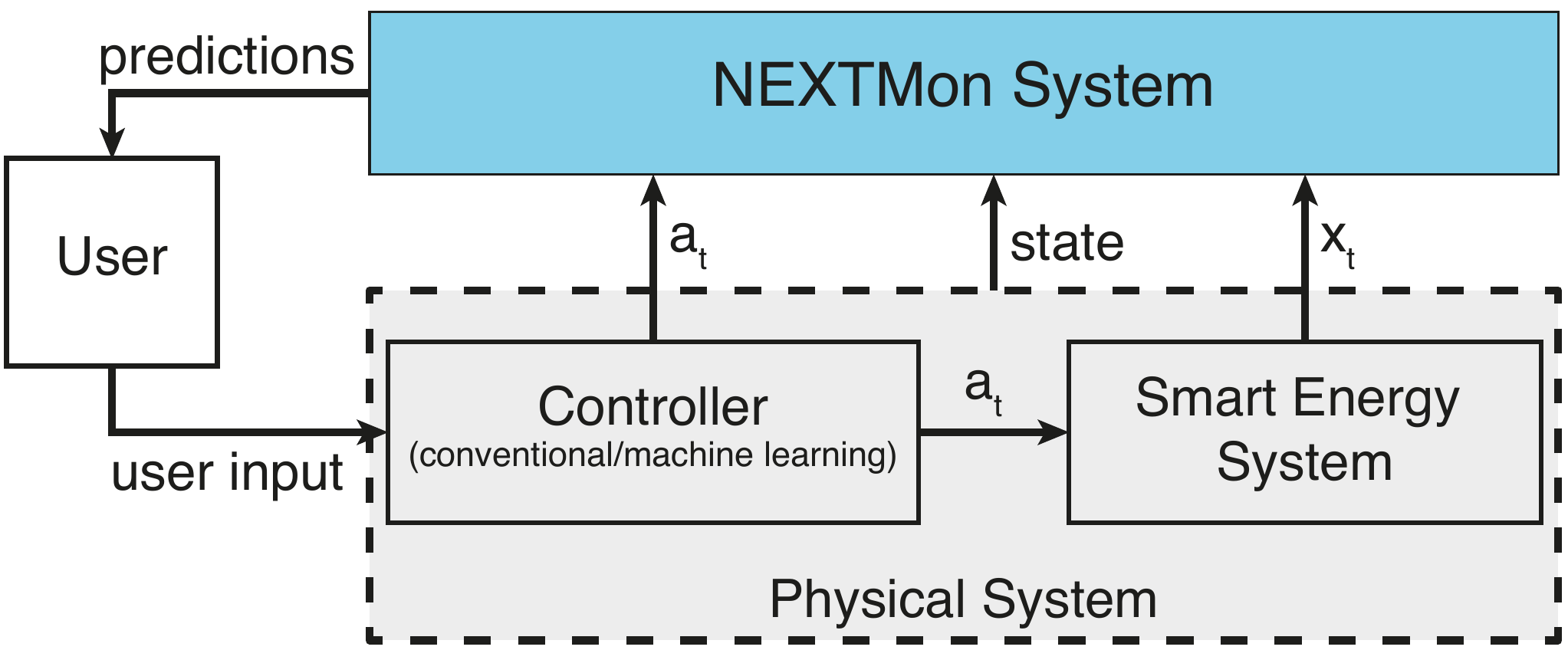}
\vspace{-1em}
\caption{A human-in-the-loop smart energy system.}
\label{fig:nextmon}
\vspace{-1em}
\end{figure}

Keeping this simple example in mind, the \textit{NEXTMon} system uses 
the nexting algorithm as a model free prediction technique for calculating predictions 
as additional information for an experienced operator. It is obvious that 
nexting cannot deliver perfect predictions about future system states and 
therefore a direct control using the prediction data could be risky. Therefore, 
we propose a system which can be used as an extension to existing monitoring 
techniques delivering short term predictions for a complex energy system. The nexting 
algorithm does not require either a complete system model nor special sensors. 
Furthermore, it can calculate predictions for several timescales without 
a drastically increase of computational power due to the limited information 
needed for updating the weight vectors and information stored in memory. 
Also the feature extraction using tile coding does not require strong expert 
knowledge since it uses just scaled versions of all available sensor signals 
and event signals. Only setting up different joint coding groups requires 
some experience in high dimensional systems. 

Installing the nexting algorithm on an existing system does not require any 
changes on the system itself. It only needs access to all available data and 
control signals (see Figure \ref{fig:nextmon}). After an initial learning phase, 
which depends on the number of unique system states, the nexting algorithm is 
not only able to predict each sensor state which was previously observed, but 
extrapolate this acquired knowledge about the system to new, unseen situations. 
Such a capability can be explained by the generalization capability of tile 
coding. Namely, states that are previously not recorded are mapped closely to 
already experienced ones and therefore sharing information about them.

Furthermore, we can use the generated predictions together with a redundant 
controller to calculate a control strategy using the predicted data. Those 
predicted control actions can be used to highlight upcoming actions which will 
be undertaken by the system currently monitored through the NEXTMon architecture. 
Those highlighted actions are potentially useful for a human operator to decide 
if a monitored system needs attention in the near future. In the following 
experiment we demonstrate a thermal house model which is heated to a 
desired temperature while the outdoor temperature varies according to natural 
temperature data of a weather station.

\begin{figure*}
  \includegraphics[width=\textwidth]{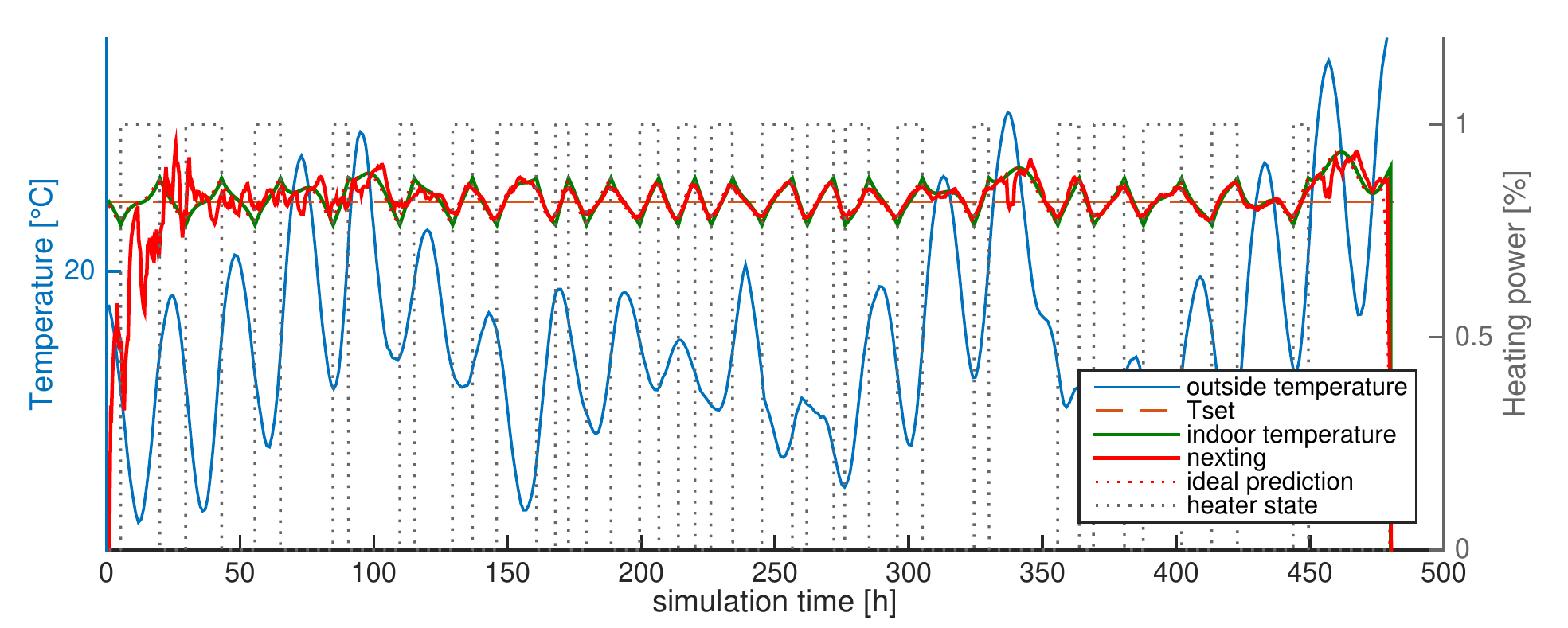}
  \caption{Comparison of simulated indoor temperature (red) data against the 
  values predicted by the nexting algorithm (green) including the effects of 
  varying outdoor temperatures (blue) and a threshold controlled heater (black). 
  The system needs about 100 hours of learning data to achieve good prediction 
  results.}
  \label{fig:next_heater}
\end{figure*}

\section{Experimental Settings}
\label{sec:experiment}
In our experiment we use a thermal house model to simulate temperature 
variations within a building. We focused on implementing primary effects 
of heat loss and heat production within a building. The differential equation 
for the indoor temperature is described by
\begin{equation}
\label{eq:deltaTindoor}
\frac{dT_{in}}{dt} = - \alpha (T_{in} - T_{out}) + \beta P_t,
\end{equation}
where $T_{in}$ is the indoor and $T_{out}$ the outdoor temperature. The 
factors $\alpha = \frac{1}{C} (a_{windows} \cdot u_{windows} + a_{walls} \cdot u_{walls})$ 
and $\beta = \frac{\eta}{C}$ describe the characteristics of the house model. 
We use a simple one-room model where the window area is given by $a_{windows} = 2m^3$ 
with a corresponding thermal transmittance $u_{windows} = 50\frac{W}{m^2K}$ 
(U-values). 
The area of all walls is $a_{walls} = 10m^3$ with a thermal transmittance of 
$u_{walls} = 1\frac{W}{m^2K}$. 
Total heat capacity is approximated by $C = C_{air} + C_{furniture} + C_{walls}$, 
where we assume that the room has a volume of $30m^3$ and $200kg$ of furniture thus 
$C_{air} = 39000\frac{J}{K}$, $C_{furniture} = 840000\frac{J}{K}$, and 
the effect of the walls contribute with $C_{walls} = 9 \cdot 10^6\frac{J}{K}$. 
The heater has a total power $P = 2000W$ and an efficiency factor of $\eta = 0.8$. 
It is assumed that the heating power $P$ together with the $\beta$ raises the 
room temperature according to the power consumed (convection and other thermal heat 
radiation is neglected). The outdoor temperature $T_{out}$ is simulated using 
20 days of hourly recorded temperature measurements \cite{energyplus2001} 
starting in April \tcg{of a weather station in Berlin}. The model was simulated 
for one minute time steps resulting in $N = 28800$ data points. 

There are certainly more sophisticated controllers like PID controllers for 
controlling the heater according to a set point. For simplicity and better 
readability of the 
selected actions, we decided to use an on-off controller with a hysteresis 
of one degree Celsius. Therefore, the heater is fully turned on if the temperature 
falls one degree below the set point and is turned off again if the temperature has 
raised to $T_{set} + 1^{\circ}C$. 

\section{Simulation Results}
\label{sec:result}
In Figure \ref{fig:next_heater} the simulation output and the prediction results 
for a prediction time horizon of 50 minutes ($\gamma = 0.98$) are depicted. The temperature 
set point $T_{set}$ was set to $23^{\circ}C$. If the heater was turned on the 
black dotted signal is set to 1 corresponding to full power ($P_t = 2000W$) and reset to 
0 after turning it off. The outdoor temperature is plotted in blue, the simulated 
indoor temperature is depicted in red overlayed by the nexting prediction (green). 
Due to the binary feature representation used for the nexting algorithm the predictions 
are affected with short peaks and can be filtered (moving average) afterwards 
to get clearer prediction results. 

Two main results are visible in Figure \ref{fig:next_heater}: In the beginning, 
the algorithm needs some amount of samples to learn the weight vector in order to 
produce usable results. After that initial phase it is is able to predict the 
indoor temperature signal also in the case, where the natural outdoor temperature 
changes. It is important to notice that the type of controller is unknown to 
the NEXTmon system. The length of the initial learning phase depends on the number 
of unique states to be observed (the learning curve of the nexting algorithm is 
currently an active research topic). 
\begin{figure}
\centering
\includegraphics[width=0.45\textwidth]{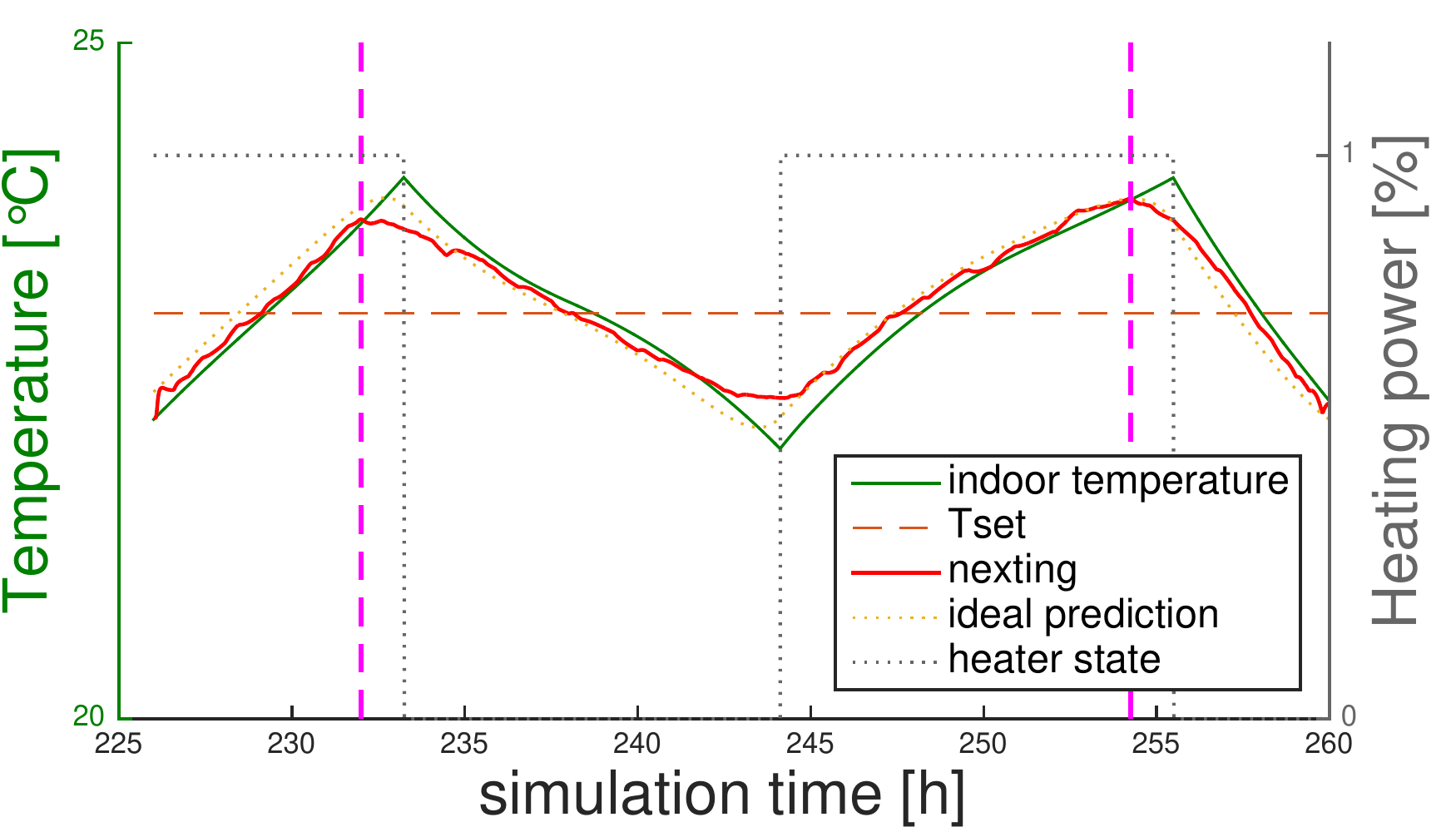}
\caption{Simulation result of a human-in-the-loop control system with predicted 
switching points.}
\label{fig:next_heater_crop}
\end{figure}
In Figure \ref{fig:next_heater_crop} we zoom in the period between 
$t = 225h$ up to $t = 260h$ in order to better visualize the resulting 
predictions. There are three different kinds of predictions. The ideal prediction 
(dotted line) offline calculated using Equation \ref{eq:predictioninmultitimescalenexting}, 
the predicted indoor temperature (red line), and the predicted actions 
outputted by a local maxima detection algorithm. The ideal prediction corresponds 
to the best prediction possible under the assumption that the full signal 
is known. This ideal prediction could be achieved theoretically by the nexting algorithm 
with an infinitely long feature vector (highest precision of tile coding) and after an 
infinitely long learning phase. In practice, the nexting prediction does not 
achieve this optimum but approximates it very well and is most of the time 
precursory to the signal to be predicted. 
Applying a local maxima detection algorithm to the predicted indoor temperature 
signals enables to detect the switching off events (c.f. Figure 
\ref{fig:next_heater_crop}, vertical dashed lines). This approach could be compared 
to the heuristic a human operator would apply during monitoring the raw predictions. 
Currently obtaining reasonable general numerical results of the NEXTmon system is
difficult. Calculating the root mean squared error (RMSE) according to the 
ideal prediction or the source signal would be possible. But we have seen that 
in different domains the RMSE is not a useful qualitative characteristic. Also 
the number of successful predicted events does not deliver a good qualitative figure. 
Its accuracy is highly depended on the local maxima detection algorithm which 
depends again on the problem.  Beyond that, our idea is the development 
of a human-in-the-loop monitoring system enriched with predicted raw sensor 
signals. Only for predictive control systems the absolute prediction accuracy 
would matter, but this was not the intended goal. 
\section{Conclusions}
\label{sec:conclusion}
Monitoring complex energy systems often require fast reactions. However, 
a complete system model or the external factors (like weather) cannot be 
modeled accurately for a prediction or simulation algorithm. Therefore, 
we designed the NEXTMon architecture, a model free prediction algorithm for raw 
sensor signals delivering predictive information while keeping the human-in-the-loop. 
In our experiments, we verified that this framework is able to supply sufficient 
information to detect upcoming actions enabling a human operator to reason about 
future system states. This could avoid faulty decisions in complex environments and 
extends the time an operator has to focus her attention to a monitored system 
before an upcoming action is executed. 

It is also obvious that the nexting algorithm can produce inaccurate results 
and does not always deliver optimal predictions that lie in a certain error threshold.
This drawback is compensated by 
the applicability of the algorithm in all systems where raw sensor signals 
are available and used for monitoring purposes. The NEXTMon system does not require 
significant changes to the energy system and can use additional external 
signals and measurements (weather forecasts, schedules) to learn an 
inherent and flexible model for the predictions. Furthermore, 
compared to other prediction algorithms the nexting algorithm is data 
efficient and needs less historic data at each update step than comparable 
algorithms. 

Current advances in integrating the NEXTMon architecture into a distributed 
wind power generation system together with weather forecasts show promising 
results. In the future, we plan to integrate the architecture in more complex 
domains and in productive scenarios. For this, additional formal analysis and 
verification of the predictions are needed. Also an ergonomic user interface 
for the online usage has to be developed.With the NEXTMon architecture we have 
proposed a framework for integrating short-term predictions into a 
human-in-the-loop monitoring and control scenario enabling proactive 
decisions of an human operator.

\bibliographystyle{IEEEtran}
\bibliography{IEEEabrv,biblio.bib}

\begin{thebibliography}{10}
\providecommand{\url}[1]{#1}
\csname url@samestyle\endcsname
\providecommand{\newblock}{\relax}
\providecommand{\bibinfo}[2]{#2}
\providecommand{\BIBentrySTDinterwordspacing}{\spaceskip=0pt\relax}
\providecommand{\BIBentryALTinterwordstretchfactor}{4}
\providecommand{\BIBentryALTinterwordspacing}{\spaceskip=\fontdimen2\font plus
\BIBentryALTinterwordstretchfactor\fontdimen3\font minus
  \fontdimen4\font\relax}
\providecommand{\BIBforeignlanguage}[2]{{%
\expandafter\ifx\csname l@#1\endcsname\relax
\typeout{** WARNING: IEEEtran.bst: No hyphenation pattern has been}%
\typeout{** loaded for the language `#1'. Using the pattern for}%
\typeout{** the default language instead.}%
\else
\language=\csname l@#1\endcsname
\fi
#2}}
\providecommand{\BIBdecl}{\relax}
\BIBdecl

\bibitem{smart2015hayes}
B.~Hayes, J.~Gruber, and M.~Prodanovic, ``A closed-loop state estimation tool
  for {MV} network monitoring and operation,'' \emph{IEEE Transactions on Smart
  Grid}, vol.~6, no.~4, pp. 2116 -- 2125, July 2015.

\bibitem{failure2011calderaro}
V.~Calderaro, C.~N. Hadjicostis, A.~Piccolo, and P.~Siano, ``Failure
  identification in smart grids based on petri net modeling,'' \emph{IEEE
  Transactions on Industrial Electronics}, vol.~58, no.~10, pp. 4613 -- 4623,
  October 2011.

\bibitem{collins2012smart}
K.~Collins, M.~Mallick, G.~Volpe, and W.~Morsi, ``Smart energy monitoring and
  management system for industrial applications,'' in \emph{IEEE Electrical
  Power and Energy Conference (EPEC)}, October 2012, pp. 92 -- 97.

\bibitem{sequeira2014energy}
H.~Sequeira, P.~Carreira, T.~Goldschmidt, and P.~Vorst, ``Energy cloud:
  Real-time cloud-native energy management system to monitor and analyze energy
  consumption in multiple industrial sites,'' in \emph{EEE/ACM 7th
  International Conference on Utility and Cloud Computing (UCC)}, December
  2014, pp. 529 -- 534.

\bibitem{linda2012comp}
O.~Linda, D.~Wijayasekara, M.~Manic, and C.~Rieger, ``Computational
  intelligence based anomaly detection for building energy management
  systems,'' in \emph{IEEE 5th International Symposium on Resilient Control
  Systems (ISRCS)}, August 2012, pp. 77 -- 82.

\bibitem{nexting2014modayil}
J.~Modayil, A.~White, and R.~S. Sutton, ``Multi-timescale nexting in a
  reinforcement learning robot,'' \emph{Adaptive Behavior}, vol.~22, no.~2, pp.
  146--160, April 2014.

\bibitem{gunt:mach15}
J.~G{\"u}nther, P.~M. Pilarski, G.~Helfrich, H.~Shen, and K.~Diepold,
  ``Intelligent laser welding through representation, prediction, and control
  learning: An architecture with deep neural networks and reinforcement
  learning,'' Accepted at Mechatronics, 2015.

\bibitem{gilani2013importance}
S.~Gilani, S.~Windmann, F.~Pethig, B.~Kroll, and O.~Niggemann, ``The importance
  of model-learning for the analysis of the energy consumption of production
  plants,'' in \emph{IEEE 18th Conference on Emerging Technologies Factory
  Automation (ETFA)}, September 2013, pp. 1 -- 8.

\bibitem{sutton1998rl}
R.~S. Sutton and A.~G. Barto, \emph{{Reinforcement Learning: An Introduction}},
  1st~ed.\hskip 1em plus 0.5em minus 0.4em\relax Cambridge, MA, USA: MIT Press,
  1998, vol.~1, no.~1.

\bibitem{energyplus2001}
\emph{International Weather for Energy Calculations (IWEC Weather Files) Users
  Manual and CD-ROM}, ASHRAE, 2001.

\end{thebibliography}

\end{document}